\documentclass[sn-mathphys]{sn-jnl}% Math and Physical Sciences Reference Style
\jyear{2021}%
\usepackage{makecell}
\theoremstyle{thmstyleone}%
\theoremstyle{thmstyletwo}%

\theoremstyle{thmstylethree}%

\begin{document}

\title[Article Title]{Universal Graph Filter Design based on Butterworth, Chebyshev and Elliptic Functions}

%%=============================================================%%
%% Prefix	-> \pfx{Dr}
%% GivenName	-> \fnm{Joergen W.}
%% Particle	-> \spfx{van der} -> surname prefix
%% FamilyName	-> \sur{Ploeg}
%% Suffix	-> \sfx{IV}
%% NatureName	-> \tanm{Poet Laureate} -> Title after name
%% Degrees	-> \dgr{MSc, PhD}
%% \author*[1,2]{\pfx{Dr} \fnm{Joergen W.} \spfx{van der} \sur{Ploeg} \sfx{IV} \tanm{Poet Laureate} 
%%                 \dgr{MSc, PhD}}\email{iauthor@gmail.com}
%%=============================================================%%

\author[1]{\fnm{Zirui} \sur{Ge}}\email{1019010430@njupt.edu.cn}

\author[1]{\fnm{Haiyan} \sur{Guo}}\email{guohy@njupt.edu.cn}
%\equalcont{These authors contributed equally to this work.}

\author[1]{\fnm{Tingting} \sur{Wang}}\email{2018010215@njupt.edu.cn}
\author*[1]{\fnm{Zhen} \sur{Yang}}\email{yangz@njupt.edu.cn}
%\equalcont{These authors contributed equally to this work.}
\affil[1]{\orgdiv{School of Communication and Information Engineering}, \orgname{ Nanjing University of Posts and Telecommunications}, \city{Nanjing}, \postcode{210023}, \country{China}}

%%==================================%%
%% sample for unstructured abstract %%
%%==================================%%

\abstract{Graph filters are crucial tools in processing the spectrum of graph signals. In this paper, we propose to design universal IIR graph filters with low computational complexity by using three kinds of functions, which are Butterworth, Chebyshev, and Elliptic functions, respectively. Specifically, inspired by the classical analog filter design method, we first derive the zeros and poles of graph frequency responses. With these zeros and poles, we construct the conjugate graph filters to design the Butterworth high pass graph filter, Chebyshev high pass graph filter, and Elliptic high pass graph filter, respectively. On this basis, we further propose to construct a desired graph filter of low pass, band pass, and band stop by mapping the parameters of the desired graph filter to those of the equivalent high pass graph filter. Furthermore, we propose to set the graph filter order given the maximum passband attenuation and the minimum stopband attenuation. Our numerical results show that the proposed graph filter design methods realize the desired frequency response more accurately than the autoregressive moving average (ARMA) graph filter design method, the linear least-squares fitting (LLS) based graph filter design method, and the Chebyshev FIR polynomial graph filter design method. }

\keywords{graph filters, Butterworth functions, Chebyshev functions, Elliptic functions}

%%\pacs[JEL Classification]{D8, H51}

%%\pacs[MSC Classification]{35A01, 65L10, 65L12, 65L20, 65L70}

\maketitle

\section{Introduction}\label{sec1}

\par Graph signal processing (GSP) extends classical digital signal processing (DSP) technologies to signals that reside on irregular structures \cite{Ref1,Ref2}. Graph filter design is an important research area of GSP. It aims to reserve the desired graph frequencies and attenuate the others. Due to its wide range of applications, graph filter design has attracted increasing attention and shown its advantage in graph wavelets \cite{Ref3,Ref4}, graph signal denoising \cite{Ref8,Ref9,Ref10}, graph signal recovery \cite{Ref11}, speech enhancement \cite{Ref5,Ref6} and others.

\par There are generally two types of graph filters, which are finite impulse response (FIR) graph filters \cite{Ref7, Ref12, Ref13, Ref14} and infinite impulse response (IIR) graph filters \cite{Ref15, Ref16, Ref17,Ref18}. FIR graph filters are generally modeled as the polynomials of the graph shift operator. In \cite{Ref7}, a linear least-squares fitting (LLS) based method was proposed to obtain the coefficients of FIR graph filters. In \cite{Ref12}, Shuman \emph{et al}. proposed a method to approximate the graph multipliers by Chebyshev polynomials. In \cite{Ref13}, Segarra et al. proposed to design optimal graph filters by using arbitrary linear transformations between graph signals. In \cite{Ref14}, Contino  \emph{et al}. proposed to extend distributed FIR graph filters to their edge-variant (EV) versions. It is worth noting that a high filter order is required so that the designed FIR graph filter can realize a sharp transition. However, these can be trapped in an ill condition if it meets a high order.
\par By contrast, IIR graph filters are generally modeled as rational functions, which outperform polynomials and have low degrees in fitting desired graph filters. In \cite{Ref15,Ref16}, the authors first proposed the concept of the IIR graph filter. In \cite{Ref17,Ref18}, the authors proposed the concept of universal graph filter which makes graph filters applicable for any graph structure, and further designed autoregressive moving average (ARMA) graph filters by utilizing classical temporal ARMA filters. In \cite{Ref25}, the authors proposed to construct the Chebyshev I graph filter by using closed-form filter coefficients, which avoids the high computational complexity led by solving complex optimization problems.
\par In this context, we investigate the universal graph filter design based on the undirected graph. Different from \cite{Ref15} and \cite{Ref25}, we propose to construct the desired Butterworth/Chebyshev/Elliptic IIR graph filter by a pair of conjugate graph filters. Specifically, we first obtain the zeros and poles of graph frequency responses based on Butterworth, Chebyshev, and Elliptic functions, respectively, and further construct the conjugate graph filters to design three types of desired graph filters.

\par The main contributions of this paper are summarized as follows.

\par 1)	Inspired by the classical Butterworth, Chebyshev, and Elliptic functions, we derive the zeros and poles of the corresponding graph frequency responses and further build the zero-pole equations which are used for the high pass universal IIR graph filter design. Furthermore, we propose to design low pass, band pass, and band stop graph filters mapped by the obtained high pass graph filters.
\par 2)	For the integrity of the graph filter design, we also propose to set the order of IIR graph filters with the given requirements.
\par 3)	Our simulation results show that our designed IIR filters perform better than the ARMA graph filter, LLS graph filter, and Chebyshev FIR polynomial graph filter in terms of graph filter responses. Additionally, the proposed Elliptic graph filter has the best performance in approximating the step graph frequency response.

\par The outline of the paper is as follows. Section II introduces the basic concepts of signal processing on graphs and classical filters. Section III shows the construction of high pass, low pass, band pass, and band stop graph filters. In Section IV, we show how to set the order of IIR graph filters. Our simulation results are provided in Section V, while Section VI concludes the paper.

\section{Related Work}\label{sec2}
\subsection{Graph Filters.} \label{sec2}
\par Consider an undirected graph $G=\left( V,E \right)$ with $N$ nodes and $M$ edges, where $
V=\left[\begin{array}{llll}
v_{1} & v_{2} & \cdots & v_{N}
\end{array}\right]
$
denotes the set of $N$  vertices and $E$ denotes the set of $M$ edges. The underlying structure of $G$ is described by the adjacency matrix $A\in R^{N\times N}$ or by Laplacian matrix $L=D-A$, where $D$ is the diagonal degree matrix. Since $G$ is undirected, the edge between $v_i$ and $v_j$ is same as that in  $v_j$ and $v_i$. It is worth noting that the Laplacian matrix $L$ is symmetric, and its eigen-decomposition can be further expressed as $L=U\varLambda U^T$, where $U$ is the eigenvector matrix and $\varLambda $ is the diagonal matrix. 
\par In \cite{Ref25, Ref26}, the authors proved that the eigenvalues of the normalized Laplacian matrix are in the interval [0, 2]. In \cite{Ref24}, the authors defined the eigenvector matrix $U$ as the graph Fourier transformation basis and the diagonal elements of the diagonal matrix $\varLambda $ as the graph frequencies $\lambda _1,\lambda _2,…,\lambda _N$. In \cite{Ref7}, the authors proposed to sort the graph frequencies by using the total variation, so that small eigenvalues represent high frequencies and large eigenvalues represent low frequencies. Therefore, a high pass graph filter aims to retain small eigenvalues and discard large eigenvalues, while a low pass graph filter aims to retain large eigenvalues and discard small eigenvalues.
\par By taking the normalized Laplacian matrix as the graph shift operator. The FIR graph filter is expressed as

$$
H=\sum_{i=0}^{N-1}{h_iL_n^i}.\eqno(1)
$$
and the IIR graph filter \cite{Ref15} is 
$$
H=\left(\sum_{i=0}^{q} h_{i} L_{n}^{i}\right)^{-1} \sum_{i=0}^{p} h_{i} L_n^i, \quad \sum_{i=0}^{q} h_{i} L_{n}^{i} \text { is invertible }.\eqno(2)
$$
\par 	The both graph filters' graph frequency responses are written as 
$$
H\left( \lambda \right) =\sum_{i=0}^{N-1}{h_i\lambda ^i},\eqno(3)
$$
and
$$
H\left( \lambda \right) =\frac{p_n\left( \lambda \right)}{p_d\left( \lambda \right)}=\frac{\sum_{p=0}^P{a_p\lambda ^p}}{\sum_{q=0}^Q{b_q\lambda ^q}},\eqno(4)
$$
\noindent respectively. $P$,$Q$ are the highest order in the numerator and denominator. It is noted that $\sum_{q=0}^Q{b_qL_{n}^{q}}$ is invertible to guarantee a bounded output given a bounded input. That is the stability of a graph filter. 
\subsection{Classical Analog IIR Filters. }\label{sec2}
\par The magnitude response of an analog IIR filter is given by
$$
\lvert H\left( \Omega \right) \rvert^2=\frac{1}{1+\varepsilon_{p}^{2}F_{N}^{2}(\omega )},\omega =\frac{\Omega}{\Omega_p} .\eqno(5)
$$
where $N$ is the filter order, $\varepsilon_p$ is the passband ripple parameters, $\Omega$ is the analog frequency, $\Omega _p$ is the passband cutoff frequency, and $F_N(\omega )$ is a function of the normalized frequency $\omega $.
\par In classical analog IIR filters, Butterworth, Chebyshev, and Elliptic filter design methods are well established and generally applied in different fields \cite{Ref19}.
\par For the Butterworth filter, $F_B(\omega )=\omega ^N$. The Butterworth filter keeps a flat frequency response in the pass band and stop band.
\par For the Chebyshev I filter, 
$$
F_{CI}(\omega )=C_N(x)=\begin{cases}
	\cos \left( N\cos ^{-1}x \right) , x\leqslant 1\\
	\mathrm{ch}\left( N\mathrm{ch}^{-1}x \right) , x>1\\
\end{cases}. \eqno(6)
$$
\par For the Chebyshev II filter, $F_{CII}(\omega )=\left[ k_1C_N\left( k^{-1}\omega ^{-1} \right) \right] ^{-1}$, where $k=\frac{\Omega _p}{\Omega _s}$, $k_1=\frac{\varepsilon _p}{\varepsilon _s}$, and $\varepsilon _s$ is the stopband ripple parameters. The main difference between the graph frequency responses of the Chebyshev I filter and Chebyshev II filter is that the former keeps equal ripple in the passband while the latter in the stopband.
\par	For the Elliptic filter, 
$$
F_E(\omega )=\mathrm{cd}\left( NuK_1,k_1 \right) ,\quad \omega =\mathrm{cd(}uK,k), \eqno(7)
$$
where $cd\left( x,k \right) $ is the Jacobian elliptic function with the modulus $k$ \cite{Ref20}, $u$ is a real number in the interval [0,1], $K$ and $K_1$ are the complete integral of modulus $k$ and $k_1$, respectively.
\par It is noted that $F(\omega )$ is normalized for all the four classical analog IIR filters stated above, that is, $F(1)=1$. Additionally, to guarantee the equal-ripple characteristic in the Elliptic case, the function of $F(\omega )$ satisfies the identity
$$
F(\omega )=\frac{1}{k_1F\left( k^{-1}\omega ^{-1} \right)}.\eqno(8)
$$
\par The key to designing analog filters is to obtain the poles and zeros of the magnitude responses. With the obtained zeros and poles, analog filters can be constructed as rational functions. 

\section{Graph Filter Design Based on Three Functions}\label{sec3}
\par In this section, we first investigate the Butterworth graph filter design, the Chebyshev graph filter design, and the Elliptic high pass graph filter design, respectively. Then we propose to design low pass, band pass, and band stop graph filters based on the high pass graph filter design method above.
\subsection{The Construction of High Pass graph Filters}\label{sec3}
\subsubsection{The Butterworth high pass graph filter}\label{sec3}
\par By replacing $\Omega $ in (3) with graph frequency $\lambda$, we can express the frequency response of the Butterworth graph filter as
$$
H_B(\lambda )=\frac{1}{1+\varepsilon _{p}^{2}\omega ^{2N}}, \omega =\frac{\lambda}{\lambda _p}.\eqno(9)
$$
where $\lambda _p$ is the passband cutoff graph frequency. We can obtain the poles of $H_B(\lambda )$ as 
$$
\lambda =\lambda _p\varepsilon _{p}^{-1/N}e^{j\frac{2m-1}{2N}\pi},m=1,2,\cdots ,2N .\eqno(10)
$$
\par We select the first $N$ poles above the real axis, i.e., $m=1,2,\cdots ,N$, to construct a graph filter expressed as
$$
G_B\left( \lambda \right) =\frac{1}{\prod_{i=1}^N{\left( 1-\lambda /\lambda _{p_i} \right)}},\eqno(11)
$$
\noindent where $G_B\left( \lambda \right) $ can attain the desired magnitude response, but its frequency response is complex-valued only works in complex signal space. We name another filter $G_{B}^{*}\left( \lambda \right) $ constructed by the other N points below the real axis, where $G_{B}^{*}\left( \lambda \right) $ is the conjugate function with $G_B\left( \lambda \right)$.
\par Thus, to eliminate the disallowed phases, we compose the conjugated filters to construct the final IIR graph filter whose frequency response is real-valued. That is different from the classical filters, which only keep the poles of the left complex plane. We can obtain frequency response of the Butterworth graph filter as
$$
\begin{array}{l}
	H_B\left( \lambda \right) =G_B\left( \lambda \right) G_{B}^{*}\left( \lambda \right)\\
	=\frac{H_0}{\prod_{i=1}^{\lfloor N/2 \rfloor}{1+2\lambda ^2imag\left( 1/\lambda _{p_i} \right) ^2+\lambda ^4abs\left( 1/\lambda _{p_i}^{4} \right)}}\\
\end{array},\eqno(12)
$$
\noindent where 
$$
H_0=\left\{ \begin{array}{c}
	\frac{1}{1-2\lambda real\left( 1/\lambda _{p_{\lceil N/2 \rceil}} \right) +\lambda ^2abs\left( 1/\lambda _{p_{\lceil N/2 \rceil}} \right) ^2}, N\,\,is\,\,odd\\
	1, N\,\,is\,\,even\\
\end{array} \right. ,\eqno(13)
$$
\noindent and we indicate by $abs\left( \cdot \right) $, $imag\left( \cdot \right) $ and $real\left( \cdot \right) $ the absolutely value, imaginary part, and real part of a complex value. We can use the convolution operation to compute polynomial multiplication in (10). For a large $N\left( N\geqslant 64 \right) $, we can use the FFT to further simplify the complexity. 
\par This graph filter (10) is the cascade form proposed in \cite{Ref15}. In the following two cases, we also take the cascade form to construct the graph filter.
\par Fig. 1 illustrates the graph Butterworth filter’s poles for $N=16$, where Fig. 1 (a) and Fig. 1 (b) show the locations of all the $2N$ poles and those of the first $N$ poles, respectively.

\begin{figure}[H]
	{	\centering
		\includegraphics[width=4in]{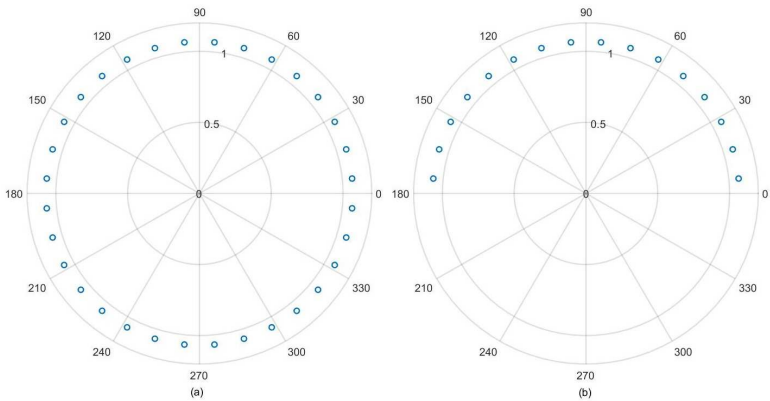}\\}
	\caption{The locations of Butterworth graph filter' poles. (a) The locations of all the 2N poles. (b) The locations of the first N poles.}
\end{figure}
\par Observe from Fig. 1 that none of the poles reside on the real axis. Thus, $\prod_{i=1}^N{\left( 1-\lambda /\lambda _{p_i} \right)}\ne 0$ is always satisfied, and the Butterworth graph filter can always provide a bounded output given a bounded input. That is, stability is guaranteed.
\subsubsection{The Chebyshev graph filter design}\label{sec3}
\par Similar to the Chebyshev I analog filter and the Chebyshev II analog filter, we can express the frequency response of the Chebyshev I graph filter and that of the Chebyshev II graph filter as
$$
H_{CI}(\lambda )=\frac{1}{1+\varepsilon _{p}^{2}C_{N}^{2}\left( \gamma \right)},\quad \gamma =\frac{\lambda}{\lambda _p}, \eqno(14)
$$
\noindent and
$$
H_{CII}(\lambda )=\frac{1}{1+\varepsilon _{p}^{2}\left[ k_{1}^{2}C_{N}^{2}\left( k^{-1}\gamma ^{-1} \right) \right] ^{-1}},\quad \gamma =\frac{\lambda}{\lambda _p}, \eqno(15)
$$
\noindent respectively.
\par We can obtain the poles of $H_{CI}(\lambda )$ as 
$$
\begin{aligned}
\lambda_{C \mathrm{I}} &=\lambda_{p} \cos \left(\frac{2 m-1}{2 N} \pi\right) \cdot \operatorname{ch}\left(\frac{1}{N} s h^{-1} \frac{1}{\varepsilon_{p}}\right) \\
& \pm j \lambda_{p} \sin \left(\frac{2 m-1}{2 N} \pi\right) \cdot \operatorname{sh}\left(\frac{1}{N} \operatorname{sh}^{-1} \frac{1}{\varepsilon_{p}}\right), m=1,2, \cdots, 2 N .
\end{aligned} .\eqno(16)
$$

\begin{figure}[H]
	{	\centering
		\includegraphics[width=4in]{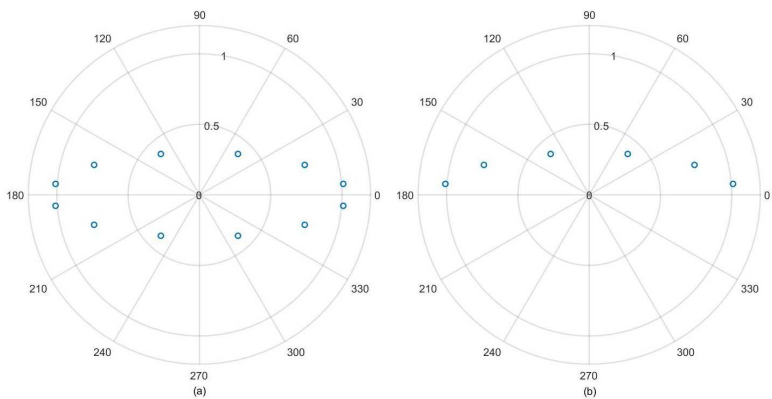}\\}
	\caption{The locations of Chebyshev I graph filter’ poles. (a) The locations of all the 2N poles. (b) The locations of the first N poles.}
\end{figure}

\par Fig. 2 illustrates the Chebyshev I graph filter’ poles in the polar coordinate for $N=6$. Similar to the Butterworth graph filter, all the $2N$ poles are selected. Observe from Fig. 2 that the poles do not reside on the real axis. The reason for this is that $sh\left( \frac{1}{N}sh^{-1}\frac{1}{\varepsilon _p} \right) \ne 0$ and $\sin \left( \frac{2m-1}{2N}\pi \right) \ne 0, m=1,2,\cdots ,2N$. Thus, the designed Chebyshev I graph filter satisfies the stability. Given any bounded input, the designed Chebyshev I graph filter always provides a limited output. Differ from \cite{Ref25}, which directly uses the poles in classical filters, we utilize the conjugate graph filters to obtain the desired graph filter in a simpler form. The construction of the Chebyshev I graph filter is the same as equation (10).

\par Let us now study the zeros and poles of $H_{CII}\left( \lambda \right) $. We can obtain the poles of $H_{CII}(\lambda )$ as 
$$
\lambda _{C\mathrm{II}}=\lambda _s\frac{c_m\pm jd_m}{c_{m}^{2}+d_{m}^{2}}, m=1,2,\cdots ,2N, \eqno(17)
$$
\noindent where $c_m=\cos \left( \frac{2m-1}{2N}\pi \right) \cdot ch\left( \frac{1}{N}sh^{-1}\varepsilon _s \right) $, $d_m=\sin \left( \frac{2m-1}{2N}\pi \right) \cdot sh\left( \frac{1}{N}sh^{-1}\varepsilon _s \right)$.
\par The zeros of $H_{CII}(\lambda )$ are 
$$
z_{CII}=\frac{\lambda _s}{\cos \frac{(2m-1)\pi}{2N}},m=1,2,\cdots ,2N.\eqno(18)
$$
\par It is worth noting that in the case where $N$ is odd, there are two zeros with infinite values since $\cos \frac{\left( 2m-1 \right) \pi}{2N}=0$, when $m=\frac{\lceil N \rceil}{2}$ and $m=\frac{\lceil N \rceil}{2}+N$, which should be discarded in the construction of graph filters.
 
\par Fig. 3 illustrates the Chebyshev II graph filter’ poles in the polar coordinate for $N=10$.

\begin{figure}[H]
	{	\centering
		\includegraphics[width=4in]{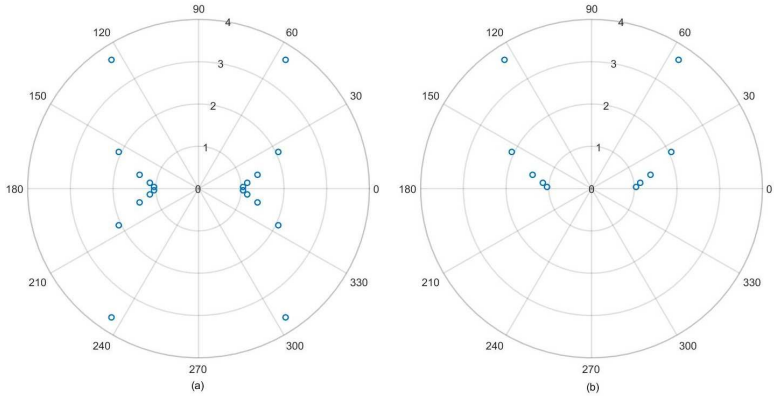}\\}
	\caption{The locations of Chebyshev II graph filter’ poles. (a) The locations of all the $2N$ poles. (b) The locations of the first $N$ poles.}
\end{figure}
\par The frequency response of the Chebyshev II graph high pass filter is expressed as 
$$
\begin{array}{l}
	H_{CII}\left( \lambda \right) =G_{CII}\left( \lambda \right) G_{CII}^{*}\left( \lambda \right)\\
	=\frac{H_0\prod_{i=1}^{\lfloor N/2 \rfloor}{1-2\lambda ^2/\lambda _{z_i}^{2}+\lambda ^4/\lambda _{z_i}^{4}}}{\prod_{i=1}^{\lfloor N/2 \rfloor}{1+2\lambda ^2imag\left( 1/\lambda _{p_i} \right) ^2+\lambda ^4abs\left( 1/\lambda _{p_i}^{4} \right)}}\\
\end{array}.\eqno(19)
$$
where $H_0$ is the same as equation (11).
\subsubsection{The Elliptic Graph Filter Design}\label{sec3}
\par We can obtain the zeros and poles of Elliptic graph filter as 
$$
z_E=\mathrm{cd(}u_mK,k), u_m=\frac{2m-1}{N},\quad m=1,2,...,2N, \eqno(20)
$$
\noindent and 
$$
\lambda _E=\lambda _p\mathrm{cd} \left( \left( u_m-jv_0 \right) K,k \right) ,\quad m=1,2,...,2N,\eqno(21)
$$
\noindent where 
$$
v_0=-\frac{j}{NK_1}\mathrm{sn}^{-1}\left( \frac{j}{\varepsilon _p},k_1 \right) .\eqno(22)
$$
\par By combining (21) and (22), we can obtain that $\lambda _E$ is complex-valued. Hence, no poles reside on the real axis so that the Elliptic graph filter is stable.
\par Fig. 4 illustrates Elliptic graph filters’ poles in the polar coordinate for $N = 8$.
\begin{figure}[H]
	{	\centering
		\includegraphics[width=4in]{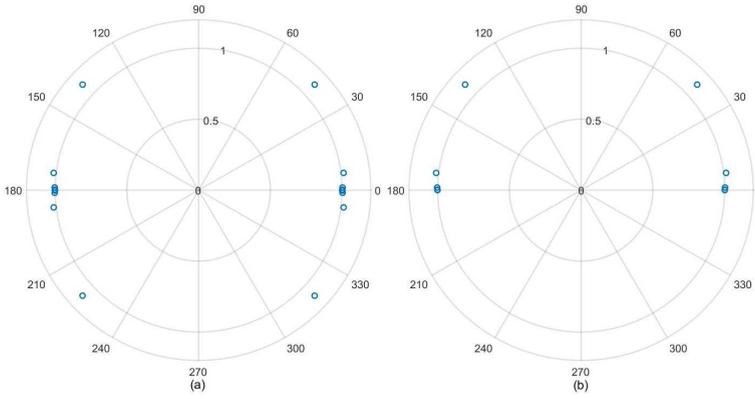}\\}
	\caption{The locations of Elliptic graph filter’ poles. (a) The locations of all the 2N poles. (b) The locations of the first N poles.}
\end{figure}
\par It is noted that the frequency response of the designed Elliptic high pass graph filter also has two forms, which are similar to that of the designed Chebyshev II graph high pass filter. Additionally, when $N$ is odd, the infinite zero points are also discarded.
\subsection{ Low pass, Band pass, and Band stop Graph filters}\label{sec3}
\par We first investigate mapping the desired low pass graph filter parameters to the equivalent high pass graph filter parameters. Let us denote the required pass band cutoff frequency and stop band cutoff frequency as $\lambda _{p}^{lp}$ and $\lambda _{s}^{lp}$, respectively. Following the parameter transformation methods in classical analog filter design \cite{Ref19}, we can obtain the graph frequency $\lambda $, the $\lambda _p$ and the $\lambda _s$ of the equivalent high pass graph filter as 
$$
\lambda =\frac{1}{\lambda ^{lp}},\quad \lambda _p=\frac{1}{\lambda _{p}^{lp}},\quad \lambda _s=\frac{1}{\lambda _{s}^{lp}} .\eqno(23)
$$
\par Let us now map the parameters of the desired band pass graph filter to those of the equivalent high pass graph filter. Upon denoting the stop band cutoff frequencies and the pass band cutoff frequencies of the desired band pass graph filter as $\lambda _{s_1}^{bp}$, $\lambda _{s_2}^{bp}$, $\lambda _{p_1}^{bp}$, $\lambda _{p_2}^{bp}$, respectively, where $\lambda _{s_1}^{bp}<\lambda _{p_1}^{bp}<\lambda _{p_2}^{bp}<\lambda _{s_2}^{bp}$. We can obtain the $\lambda $, $\lambda _p$, and $\lambda _s$ of the equivalent high pass graph filter as
$$
\lambda =\lambda ^{bp}-\frac{\lambda _{p_1}^{bp}\lambda _{p_2}^{bp}}{\lambda ^{bp}},\lambda _p=\lambda _{p_2}^{bp}-\lambda _{p_1}^{bp},\lambda _s=\min \left( \lvert \lambda _{s_1}^{bp}-\frac{\lambda _{p_1}^{bp}\lambda _{p_2}^{bp}}{\lambda _{s_1}^{bp}} \rvert,\lvert \lambda _{s_2}^{bp}-\frac{\lambda _{p_1}^{bp}\lambda _{p_2}^{bp}}{\lambda _{s_2}^{bp}} \rvert \right) .\eqno(24)
$$
\par Now we map the parameters of the desired band stop graph filters to those of the equivalent high pass graph filters. We denote the required pass band cutoff frequency and stop band cutoff frequency as $\lambda _{s_1}^{bp}$, $\lambda _{s_2}^{bp}$, $\lambda _{p_1}^{bp}$, $\lambda _{p_2}^{bp}$, respectively, where $\lambda _{s_1}^{bp}<\lambda _{p_1}^{bp}<\lambda _{p_2}^{bp}<\lambda _{s_2}^{bp}$. We can obtain the $\lambda $, $\lambda _p$, and $\lambda _s$ of the equivalent high pass graph filter as
$$
\lambda =\frac{1}{\lambda ^{bp}-\frac{\lambda _{s_1}^{bp}\lambda _{s_2}^{bp}}{\lambda ^{bp}}},\lambda _p=\max \left( \lvert \lambda _{p_1}^{bp}-\frac{\lambda _{s_1}^{bp}\lambda _{s_2}^{bp}}{\lambda _{p_1}^{bp}} \rvert,\lvert \lambda _{p_2}^{bp}-\frac{\lambda _{s_1}^{bp}\lambda _{s_2}^{bp}}{\lambda _{p_2}^{bp}} \rvert \right) ,\lambda _s=\frac{1}{\lambda _{s_2}^{bp}-\lambda _{s_1}^{bp}} .\eqno(25)
$$
\par We show an example and illustrate the correspondence between these three types of graph filters and their equivalent high pass parameters in Table 1.

\begin{table}[htbp] 
	\centering   
	\caption{Desired graph filter parameters and their Equivalent high pass parameters}  
	\begin{tabular}{lccc}      
	\hline & Low pass & Band pass & Stop pass \\  
	\hline \makecell[c]{Filter\\Parameters} & \makecell[c]{$\lambda _p=0.5$\\$\lambda _s=0.51$} & \makecell[c]{$\lambda _{s_1}=0.69$, $\lambda _{p_1}=0.7$\\$\lambda _{p_2}=1.2$, $\lambda _{s_2}=1.21$} & \makecell[c]{$\lambda _{p_1}=0.29$, $\lambda _{s_1}=0.3$\\$\lambda _{s_2}=1.7$, $\lambda _{p_2}=1.71$} \\  
	\hline 
	\makecell[c]{Equivalent high \\pass parameters} & \makecell[c]{$\lambda _{p}^{'}=0.662$\\$\lambda _{s}^{'}=0.667$} & \makecell[c]{$			\lambda _{p}^{'}=0.5$\\$\lambda _{s}^{'}=0.51$} & \makecell[c]{$\lambda _{p}^{'}=0.708$\\$\lambda _{s}^{'}=0.714$} \\
	\hline                               
	\end{tabular}

\end{table}

\section{The Graph Filter Order}\label{sec4}
\par In this section, we study how to set the order of the Butterworth, the Chebyshev I, the Chebyshev II and the Elliptic high pass graph filter, respectively.
\par Similar to analog filters, given the constraints of the passband $\lambda _p$ and the stopband $\lambda _s$, the maximum passband attenuation $R_p$ and the minimum stopband attenuation $A_s$ satisfy  
$$
-20\log \left( H\left( \lambda _p \right) /H\left( 0 \right) \right) \leqslant R_p, \eqno(26)
$$
\noindent and 
$$
-20\log \left( H\left( \lambda _s \right) /H\left( 0 \right) \right) \leqslant A_s. \eqno(27)
$$
\noindent respectively. For the Butterworth graph high pass filter, by combing (7), (26) and (27), we can obtain that 
$$
N\geqslant \log \left( \sqrt{\frac{10^{A_s/20}-1}{10^{R_p/20}-1}} \right) /\log \left( \frac{\lambda _s}{\lambda _p} \right) . \eqno(28)
$$
\par By similar analysis, we can conclude that for the Chebyshev I graph high pass filters, $N$ satisfies
$$
N\geqslant ch^{-1}\left( \sqrt{\frac{10^{A_s/20}-1}{10^{R_p/20}-1}} \right) /ch^{-1}\left( \frac{\lambda _s}{\lambda _p} \right) .\eqno(29)
$$
\par It is noted that for the Chebyshev II graph filter, $k$ should be adjusted to guarantee that the graph filter is equal-ripple. The adjusted $k$ is given by
$$
k=1/ch\left( ch^{-1}\left( \frac{1}{k_1} \right) /N \right) . \eqno(30)
$$
\par Similarly, for the high pass Elliptic graph filter, 
$$
N\geqslant \frac{K\cdot K_{1}^{'}}{K'\cdot K_1},k_1=\sqrt{\frac{10^{A_s/20}-1}{10^{R_p/20}-1}}. \eqno(31)
$$
\par It is worth noting that $K_{1}^{'}$ is the complete integral with $k_{1}^{'}$ and $k_{1}^{'}=\sqrt{1-k_{1}^{2}}$. Additionally, the parameter $k$ also needs to be adjusted as \cite{Ref20}
$$
k=\sqrt{1-\left( k' \right) ^2}, \eqno(32)
$$
\noindent and $k^{'}=\left( k_{1}^{'} \right) ^N\prod_{i=1}^L{\mathrm{s}n^4}\left( u_iK_{1}^{'},k_{1}^{'} \right) , $ where $L=\lfloor \frac{N}{2} \rfloor $.

\section{Simulation Results}\label{sec5}

\par In this section, we present our numerical results of the proposed methods and compare them with the other graph filters. The ARMA graph filter in \cite{Ref17,Ref18}, the LLS graph filter in \cite{Ref7}, as well as the FIR Chebyshev graph filter in [12] are used as the benchmarks. Without special explanation, the passband cutoff frequency is set to $\lambda _p=1$ and the stopband cutoff frequency is set to $\lambda _s=1.2$. The maximum attenuation of the passband and minimum attenuation of the stopband are set to $R_p$=1dB and $A_s$=30dB, respectively.

\begin{figure}[H]
	{	\centering
		\includegraphics[width=4.5in]{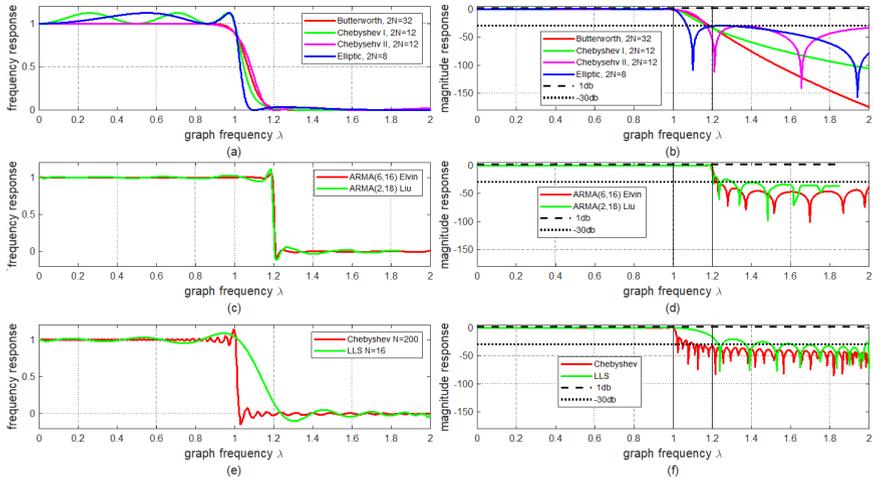}\\}
	\caption{Graph frequency responses versus graph frequency of the proposed universal IIR graph filters and the benchmarks. (a) Graph frequency response versus graph frequency of the proposed Butterworth high pass graph filter, of the proposed Chebyshev I high pass graph filter, of the proposed Chebyshev II high pass graph filter, and of the proposed Elliptic high pass graph filter. (b) Graph frequency response versus graph frequency of the ARMA graph filter. (c) Graph frequency response versus graph frequency of the LLS graph filter and the FIR Chebyshev graph filter. (b), (d), (f) are the magnitude responses of (a), (c) and (e), respectively.}
\end{figure}

\par In Fig. 5, we show the graph frequency responses versus the graph frequency of the proposed Butterworth high pass graph filter, of the proposed Chebyshev I high pass graph filter, of the proposed Chebyshev II high pass graph filter, of the proposed high pass Elliptic graph filter, of the ARMA high pass graph filter, of the LLS high pass graph filter and the FIR Chebyshev high pass graph filter. Observe from Fig. 5(b), (d), (f) that our graph filters can obtain the required frequency responses, while ARMA graph filters fail in the stopband attenuation and LLS cannot achieve the desired transition passband and stopband attenuation. Additionally, in Fig. 5(e), we can see that the needed order of the Chebyshev FIR graph filter is higher than those of our graph filters, and its fluctuations at cutoff frequency remain constant with an increasing order.

\par Fig. 6 shows graph frequency responses versus graph frequency and graph magnitude responses versus graph frequency of the proposed Butterworth graph filter, of the proposed Chebyshev I graph filter, of the proposed Chebyshev II graph filter, and of the proposed Elliptic graph filter, for $\lambda _p=1$, $\lambda _s=1.01$, $R_p$=0.1dB and $A_s$=40dB. From Fig. 6, we can see that the proposed IIR graph filters can approximate ideal graph filters if the filter order is high enough. Observe from Fig. 6 that the proposed Butterworth graph filter, the proposed Chebyshev I graph filter, the proposed Chebyshev II graph filter, and the proposed Elliptic graph filter are approximately ideal graph filters when the filter order is 910, 74, 74, and 18, respectively. The proposed Elliptic graph filter outperforms the other three filters since it can realize the desired response with the minimum order.

\begin{figure}[H]
	{	\centering
		\includegraphics[width=3.5in]{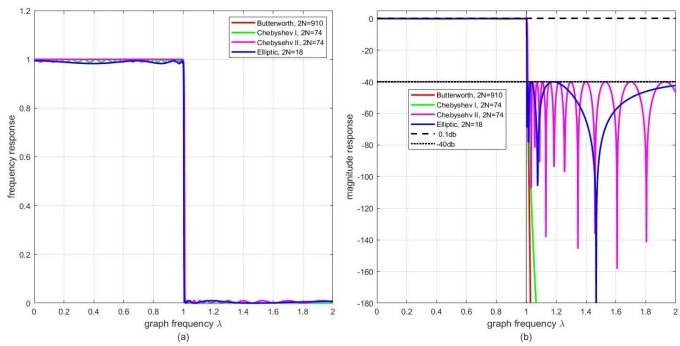}\\}
	\caption{The Graph filters frequency response versus the proposed universal IIR graph filters. (a) Graph frequency responses of three methods. (b) Magnitude responses of our three graph filters.}
\end{figure}
\par Fig. 7 shows the graph frequency responses versus the graph frequency of low pass Elliptic graph filter, band pass Elliptic graph filter, and band stop Elliptic graph filter. The parameters of three graph filters and those of their equivalent high pass graph filters are shown in Table 1. Observe from Fig. 7 that the desired low pass Elliptic graph filter, band pass Elliptic graph filter, and band stop Elliptic graph filter are successfully obtained by their corresponding equivalent high pass graph filter.
\begin{figure}[H]
	{	\centering
		\includegraphics[width=2.5in]{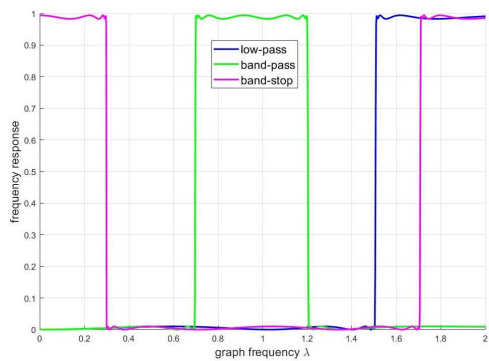}\\}
	\caption{The Graph filter frequency responses versus the proposed universal IIR graph filters. (a) Graph frequency responses of three methods. (b) Magnitude responses of our three graph filters.}
\end{figure}

\section{Conclusion}\label{sec6}
\par In this paper, we use Butterworth, Chebyshev, and Elliptic functions to design graph filters. We derive the corresponding zero-pole equations on graph frequency responses. By using these zeros and poles, we construct a couple of conjugate graph filters that achieve desired magnitude graph responses to design the universal Butterworth high pass IIR graph filters, Chebyshev high pass IIR graph filters, and Elliptic high pass IIR graph filters, respectively. We also propose to design the low pass graph filter, band pass graph filter, and band stop graph filter by mapping the parameters to those of the equivalent high pass graph filters. Furthermore, we propose to set the graph filter order given the corresponding requirements. Our simulation results show that the proposed graph filter design methods outperform the compared graph filter design methods in realizing desired frequency responses. Moreover, our methods can obtain arbitrary precision for step graph spectral responses. Additionally, the Elliptic graph filter has the minimum order for the same given requirements.
\section*{Statements and Declarations}
The authors declare that they have no known competing financial interests or personal relationships that could have appeared to influence the work reported in this paper.
\section*{Data availability}
The datasets generated during and/or analysed during the current study are available from the corresponding author on reasonable request.
\section*{Acknowledgments}
This work was supported in part by the National Natural Science Foundation of China under Grant 62071242.

\section*{References}
\bibliography{mybibfile}

\end{document}